\documentclass[twocolumn,showpacs,10pt]{revtex4-1} 

\usepackage{amsmath}
\usepackage{amssymb}
\usepackage{graphicx}
\usepackage{textcomp} 

\usepackage{amssymb}
\usepackage{xspace}
\usepackage{amsmath}
\usepackage{graphicx}
\usepackage[unicode=true,pdfusetitle, bookmarks=false, breaklinks=false,pdfborder={0 0 0},backref=false,colorlinks=false] {hyperref}

\newcommand{\rmi}{\operatorname{i}}

\usepackage{color} 
\usepackage{soul} 

\newcommand{\appropto}{\mathrel{\vcenter{
  \offinterlineskip\halign{\hfil$##$\cr
    \propto\cr\noalign{\kern2pt}\sim\cr\noalign{\kern-2pt}}}}}

\begin{document}

\title{Eigenpolarizations for Giant Transverse Optical Beam Shifts}
\date{\today}

\author{J\"org B. G\"otte}
\affiliation{Max Planck Institute for the Physics of Complex Systems, N\"othnitzer Stra{\ss}e 38, 01187 Dresden, Germany}

\author{Wolfgang L\"offler}
\affiliation{Huygens Laboratory, Leiden University, PO Box 9504, 2300 RA Leiden, The Netherlands}

\author{Mark R. Dennis}
\affiliation{HH Wills Physics Laboratory, University of Bristol, Tyndall Avenue, Bristol BS8 1TL, UK}

\begin{abstract}
We show how careful control of the incident polarization of a light beam close to the Brewster angle gives a giant transverse spatial shift on reflection.
This resolves the long-standing puzzle of why such beam shifts transverse to the incident plane [Imbert-Fedorov (IF) shifts] tend to be an order of magnitude smaller than the related Goos-H\"anchen (GH) shifts in the longitudinal direction, which are largest close to critical incidence.
We demonstrate that with the proper initial polarization the transverse displacements can be equally large, which we confirm experimentally near Brewster incidence.
In contrast to the established understanding these polarizations are elliptical and angle-dependent.
We explain the magnitude of the IF shift by an analogous change of the symmetry properties for the reflection operators as compared to the GH shift.
\end{abstract}

\pacs{11.30.Er, 42.25.Gy, 42.25.Ja}



\maketitle
\begin{figure}
\begin{center}
\includegraphics[width=\linewidth]{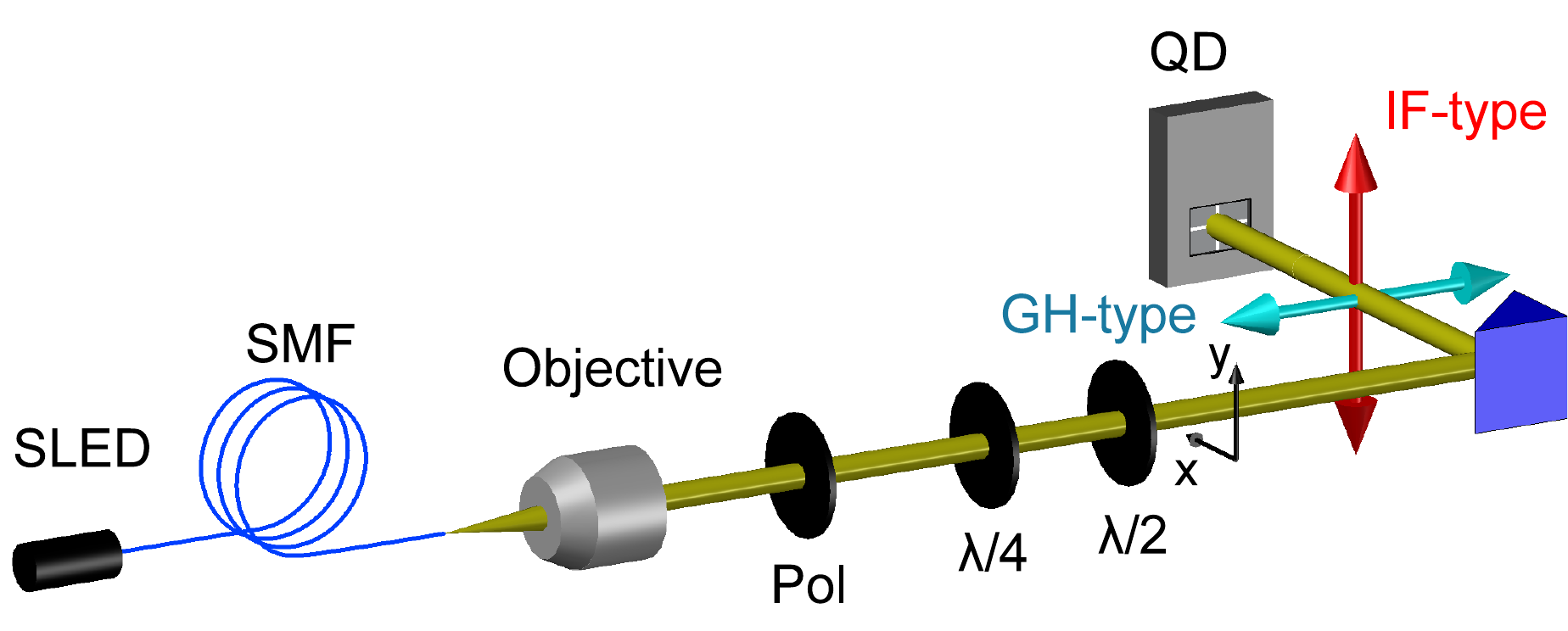}
\caption{\label{fig:expt} (Color online) Experimental setup: A collimated light beam from a single-mode fiber (SMF) coupled superluminescent diode (SLED) is prepared in each of the illuminating polarization states $\boldsymbol{f_\pm}$ alternately, using a polarizer and wave plates. After reflection at the air-glass interface, its transverse position is determined with a quadrant detector (QD).}
 \end{center}  
\end{figure}

The reflection of a plane wave from a planar isotropic interface singles out two eigenpolarizations, linear and parallel ($p$) or orthogonal ($s$) to the plane of incidence, which remain unchanged after reflection. 
Realistic optical beams, however, consist of a bundle of plane waves and, on reflection, each component experiences a slightly different reflection coefficient \cite{Artmann:AnnPhysik437:1948,BliokhAiello:JOpt15:2013}.
Accounting for these differences leads to diffractive corrections which result in a shift to the beam when compared to specular reflection.
The largest of these effects is the Goos-H\"anchen (GH) shift within the plane of incidence (see Fig.~\ref{fig:expt})  \cite{GoosHaenchen:AnnPhysik6:333,Artmann:AnnPhysik437:1948,OrnigottiAiello:JO15:2013}, for $p$ polarization and under total internal reflection.
All other polarizations result in a reduced shift \cite{AielloWoerdman:OL33:2008}. 
Beam shifts highlight many interesting aspects of optical beam physics, such as paraxiality \cite{DennisGoette:JOpt15:2013}, optical angular momentum and vortices \cite{Fedoseyev:OC193:2001,Loeffler+:PRL109:2012,DennisGoette:PRL109:2012}, and provide a classical wave analog to quantum weak measurements, including weak values \cite{Aharonov+:PRL60:1988,GoetteDennis:NJP14:2012}. 
We present here a new analysis and experiment on beam eigenpolarizations, which puts the resulting in-plane and transverse beam shifts on an equal footing.

The transverse Imbert-Fedorov (IF) shift \cite{*[] [{. Translation in Journal of Optics \textbf{15} 014002 (2013). }] Fedorov:DAN105:1955} or optical spin-Hall effect \cite{HostenKwiat:Science319:2008}, was hitherto found to be one order of magnitude smaller, reaching its maximum value under total internal reflection for circular polarizations. 
Here we find, by a careful consideration of the polarization of realistic beams on reflection, a transverse shift of similarly large magnitude as the GH shift close to critical incidence. 
Surprisingly, this shift occurs close to the Brewster angle under partial reflection, but remains purely spatial. 
We measure a differential IF shift of up to $10\,\lambda$ (8~\textmu m) which is more than $10\times$ larger compared to previous IF shift measurements (up to $\sim 0.6\,\lambda$ \cite{Pillon+:AO43:2004}).
By comparison, the GH shift measurements in the optical regime tend to about $\sim 10$ \textmu m \cite{Gilles+:OL27:2002} (this is a differential measurement of s and p shifts of respectively $25\,\lambda$ and $11\,\lambda$).

Of course, the polarization of reflected plane waves is unchanged on reflection only for $s$ and $p$ polarizations.
This eigenpolarization concept can be generalized to beams of finite width, incorporating plane waves with nearby wavevectors in the Fourier superposition.
To first order (as usual in beam shift physics), a systematic expansion of the reflection operator in Fourier space gives $\mathbf{R}(\boldsymbol{K}) = \overline{\mathbf{R}} + \boldsymbol{K} \cdot \nabla_{\boldsymbol{K}} \mathbf{R} + O(K^2)$, where $\boldsymbol{K}=(k_x, k_y)$ is the transverse wavevector.
A paraxial reflected beam is a superposition of eigenvectors of the scattering matrix based on this expansion --- we call these {\em eigenpolarizations}. As described below, for an incident eigenpolarization the shift is proportional to the eigenvalues.
Despite the fact that initially polarized beams emerge after reflection in general with an inhomogeneous polarization structure even in the paraxial regime \cite{BliokhBliokh:PRL96:2006}, this is not the case for paraxial beams initially in an eigenpolarization.
Furthermore, the linearity of reflection ensures that the shift is extremal at each of the two eigenpolarizations, suggesting these give the largest shift.
This is true even in the case, relevant in our work, when the resulting matrix is not Hermitian.

To measure the beam shift (Fig.~\ref{fig:expt}), light from a singlemode fiber-coupled superluminescent diode (SLED, 2~mW, $825 \pm 7$~nm) is collimated with a $10\times$, 0.25~NA microscope objective to obtain a beam waist of $w_0=1.8$~mm after 1~m propagation. This beam is linearly polarized and send through highest quality, zero-order $\lambda/2$ and $\lambda/4$ waveplates to prepare the incident beam in the desired eigenpolarization. To observe the beam shift, the beam is reflected from a BK7 wedge ($n=1.5103$ at $826$~nm) mounted on a rotation stage to control the angle of incidence. The position of the reflected beam is detected using a quadrant detector with a lock-in amplifier that is synchronized to the SLED modulation at 9.9~kHz.

Theoretically, beam shifts are most simply understood in Fourier space using a Jones matrix formalism in the $s,p$ basis, with the shifts being considered with respect to a `virtual reflected beam' \cite{GoetteDennis:NJP14:2012} propagating on the $z$-axis centered at the origin of the $x,y$ coordinates.
We write the transverse virtual reflected beam $\varphi(\boldsymbol{r}) \overline{\mathbf{R}}\cdot{\boldsymbol{E}},$ where $\boldsymbol{r} = (x,y),$ $\varphi(\boldsymbol{r})$ is the radially-symmetric spatial profile of the incident beam, $\boldsymbol{E}$ is the incident polarization and $\overline{\mathbf{R}} = \left(\begin{smallmatrix} -r_p & 0 \\ 0 & r_s \end{smallmatrix}\right),$ the mean reflection matrix experienced by a plane wave reflected along the $z$-direction.
In beam shift physics the plane wave reflection is the lowest order of a systematic series of correction terms which are all evaluated at the central incidence angle of the beam.
For a narrow, collimated paraxial beam with a tight spectrum in Fourier space, the full wavevector $\boldsymbol{k}$-dependent reflection matrix can be approximated by a Taylor expansion
\begin{eqnarray}
   \mathbf{R}(\boldsymbol{k}) & = & \overline{\mathbf{R}} + k_x \mathbf{R}_x + k_y \mathbf{R}_y + O(K^2), 
   \label{eq:Rexp} \\
   & = & (\mathbf{1} - \rmi k_x \mathbf{A}_x - \rmi k_y \mathbf{A}_y)\overline{\mathbf{R}} + O(K^2), 
   \label{eq:Aexp} \\
   & = & \exp(-\rmi [k_x \mathbf{A}_x + k_y \mathbf{A}_y])\overline{\mathbf{R}} + O(K^2),
   \label{eqexpA}
\end{eqnarray}
which defines the operators responsible for the shifts, referred to as `Artmann operators' in \cite{DennisGoette:NJP14:2012},
\begin{eqnarray}
   \mathbf{A}_x  & = & \frac{\rmi}{k} \left(\begin{smallmatrix} r_p'/r_p & 0 \\ 0 & r_s'/r_s \end{smallmatrix}\right),
   \label{eq:Axdef} \\
   \mathbf{A}_y  & = & \rmi \frac{\cot\theta}{k} \left( \begin{smallmatrix} 0 & -(1 + r_p / r_s) \\ (1 + r_s/r_p) & 0 \end{smallmatrix} \right),
   \label{eq:Aydef}   
\end{eqnarray}
where the reflection coefficients $r_s, r_p$ and their derivatives $r_s', r_p'$ depend on the mean incidence angle of the beam $\theta$. 
The form of Eq.~(\ref{eq:Aexp}) shows that the effect of reflection, leaving aside the mean reflection $\overline{\mathbf{R}},$ resembles a weak interaction Hamiltonian `measuring' the operators $\mathbf{A}_x, \mathbf{A}_y,$ acting on the polarization degrees of freedom \cite{Josza:PRA76:2007}, in terms of the spatial degrees of freedom $x, y$ (for which $k_x, k_y$ can be viewed as generators of translation).
This holds whenever the beam is paraxial regardless of whether the reflected light is passed through an analyzer, postselecting a single polarization component, or measuring the shift in the overall intensity \cite{GoetteDennis:NJP14:2012,DennisGoette:NJP14:2012}.

The overall shift of a beam with incident polarization $\boldsymbol{E}$ is $k^{-1}\boldsymbol{E}_{\mathrm{i}}^*\cdot\mathbf{A}_m\cdot\boldsymbol{E}_{\mathrm{i}},$ for $m = x,y,$ with $\boldsymbol{E}_{\mathrm{i}} = \overline{\mathbf{R}}\cdot\boldsymbol{E}$ is proportional to the \emph{expectation value} of $\mathbf{A}_m$ and so extremal for the eigenpolarizations, which are $s,p$ polarizations for $\mathbf{A}_x$ (GH shift) and more complicated for $\mathbf{A}_y$ (IF shift).
The shift is spatial or angular depending as to whether the eigenvalues of $\mathbf{A}_m$ are real or imaginary (usually corresponding to total or partial reflection).
On the other hand, \emph{weak values} (not measured here) correspond to postselection with analyzer polarization $\boldsymbol{F},$ giving shifts $\boldsymbol{F}^*\cdot\mathbf{A}_m\cdot\boldsymbol{E}_{\mathrm{i}}$ which are typically complex-valued, reflecting the fact that polarized components of reflected light beams are both spatially and angularly shifted \cite{GoetteDennis:NJP14:2012}.

The eigenvectors of $\mathbf{A}_x$ are clearly always given by linear $s,p$ polarizations; when reflection is total and the reflection coefficients are unimodular complex numbers, the spectrum of $\mathbf{A}_x$ is real (corresponding to the spatial GH shift), whereas otherwise, when reflection is partial, $\mathbf{A}_x$ is non-Hermitian with imaginary spectrum (corresponding to the angular GH shift) \cite{Toeppel+:NJP15:2013}.

In contrast, $\mathbf{A}_y$ is never diagonal in the $s,p$ basis, reflecting the spin-orbit origin of this term.
When reflection is total, the matrix is nevertheless Hermitian with circularly polarized eigenvectors, and eigenvalues $\pm 2 \cot \theta \cos(\arg [r_s/r_p]).$
In the regime of partial reflection, $r_p \neq r_s$ always, so the transverse Artmann matrix $\mathbf{A}_y$ is never Hermitian (nor is $\mathbf{A}_x$).
Most of the simple properties of matrices, such as the guarantee of real eigenvalues, the existence of eigenvectors, and their orthogonality, do not necessarily hold for non-Hermitian matrices.
For this reason the transverse shift in the partial regime is usually treated as ugly -- shifts are typically angular with a deformation of the underlying beam profile in both Fourier and real space.

However, $\mathbf{A}_y$ in fact always has eigenvalues, corresponding to the IF shifts
\begin{equation}
   d_{\pm} = \pm \frac{\cot\theta}{k} \left(\sqrt{\frac{r_p}{r_s}} + \sqrt{\frac{r_s}{r_p}}\right).
   \label{eq:Ayevals}
\end{equation}
Above the Brewster angle ($\theta > \theta_{\mathrm{B}})$, the ratio $r_p/r_s$ is positive and hence $d_{\pm}$ is real, implying a purely spatial shift.
Approaching the Brewster angle (from above) the shift diverges as $d_{\pm} \to \pm \infty$.
This suggests that, within the approximation (\ref{eq:Aexp}), arbitrarily large spatial shifts occur in the transverse direction arbitrarily close to the Brewster angle, in perfect analogy with the GH effect at the critical angle \cite{Gilles+:OL27:2002}.
Unlike for weak values, the first order approximation in (\ref{eq:Aexp}) retains its validity arbitrarily close to the Brewster angle, as long as the angular spread is sufficiently small, ensuring that most of the spectrum is incident at angles larger than the Brewster angle.
It is well established that large angular shifts occur at the Brewster angle \cite{Berry:PRSLA467:2011}, but not to our knowledge that spatial shifts occur {\em beyond} the Brewster angle.
A regularization of the beams shift formulas by higher orders, as for weak values \cite{Kong+:APL100:2012,DiLorenzo:PRA85:2012,GoetteDennis:OL38:2013}, would depend on the beam profile and lead to a small correction in the eigenvectors. 

As $\mathbf{A}_y$ is non-Hermitian, we have to distinguish between left and right eigenvectors in the regime of partial reflection.
Both eigenvectors correspond to elliptical polarization and change with the angle of incidence.
To observe the large IF shifts we need to act on $\mathbf{A}_y$ with its right eigenvectors, associated with the initial polarization, whereas the left eigenvectors describe final polarizations, accessible with an analyzer.
The right eigenpolarizations, corresponding to the eigenvalues (\ref{eq:Ayevals}), are
\begin{equation}
   \boldsymbol{e}_{\pm} = \frac{1}{\sqrt{|r_p+r_s|}}\left( \sqrt{|r_p|},\pm \rmi \sqrt{|r_s|}\right),
   \label{eq:evect}
\end{equation}
which are not orthogonal as $\mathbf{A}_y$ is not Hermitian.
Instead, each is orthogonal to the other left eigenvector, 
\begin{equation}
   \boldsymbol{f}_{\pm} = \frac{1}{\sqrt{|r_p+r_s|}}\left( \sqrt{|r_s|},\pm \rmi \sqrt{|r_p|}\right),
   \label{eq:levect}
\end{equation}
so $\boldsymbol{f}^\ast_{\mp} \cdot \boldsymbol{e}^{\hphantom{\ast}}_{\pm} = 0.$
For future convenience, the left eigenpolarizations $\boldsymbol{f}_\pm$ are given here as normalized to themselves rather than orthonormal to the right eigenvectors.

Clearly, when $\theta > \theta_{\mathrm{B}},$ the right and left eigenpolarizations are elliptical, with axes aligned in the $s$ and $p$ directions, and the handedness of the ellipse is given by the sign of $\pm\rmi$ in Eqs.~(\ref{eq:evect}) and (\ref{eq:levect}).
Thus each pair of eigenvectors corresponds to two identical, but oppositely oriented ellipses.
The opposite pair for the left eigenvectors are the same ellipses rotated by 90$^\circ$ (axes interchanged).
As the incidence angle approaches $\theta_{\mathrm{B}}$ (from above), the right eigenpolarizations approach pure linear $s$-polarization.
However, the right eigenpolarizations of $\mathbf{A}_y$ are not quite the incident polarizations prepared by the polarizer and the waveplates in Fig.~\ref{fig:expt}.
From the form of (\ref{eq:Aexp}), we see that the mean reflection matrix $\overline{\mathbf{R}}$ acts on an initial polarization $\boldsymbol{E}$ before $\mathbf{A}_y$ does (which is why we take expectation values with respect to $\boldsymbol{E}_{\mathrm{i}}$ and not $\boldsymbol{E}$).
In the regime $\theta > \theta_{\mathrm{B}}$ the illuminating polarizations, which yield the large spatial shifts we want to observe, are consequently given by the left eigenpolarizations $\boldsymbol{f}_\mp = \sqrt{|r_p||r_s|} \, \overline{\mathbf{R}}^{-1} \cdot \boldsymbol{e}_\pm$ as normalized vectors of the incident polarization $\overline{\mathbf{R}}^{-1} \cdot \boldsymbol{e}_\pm$ prepared by the sequence of polarizer, $\lambda/2$ and $\lambda/4$ waveplates in Fig.~\ref{fig:expt}.
From the properties of the left and right eigenpolarizations it is clear that the major axis of the illuminating polarizations $\boldsymbol{f}_{\pm}$ is rotated by 90$^\circ$ and therefore along the $p$-direction.
The distinction between the true eigenpolarization and the illuminating polarization is necessary as beam shifts can be seen as corrections to geometrical optics:
In Eq.~(\ref{eq:Rexp}), the first term $\mathbf{\overline{R}}$ acts on the polarization of an incident plane wave, while the higher terms are the corrections for a paraxial light beam that arise operationally after the plane wave reflection.

\begin{figure}
  \begin{center}
   \includegraphics[width=\linewidth]{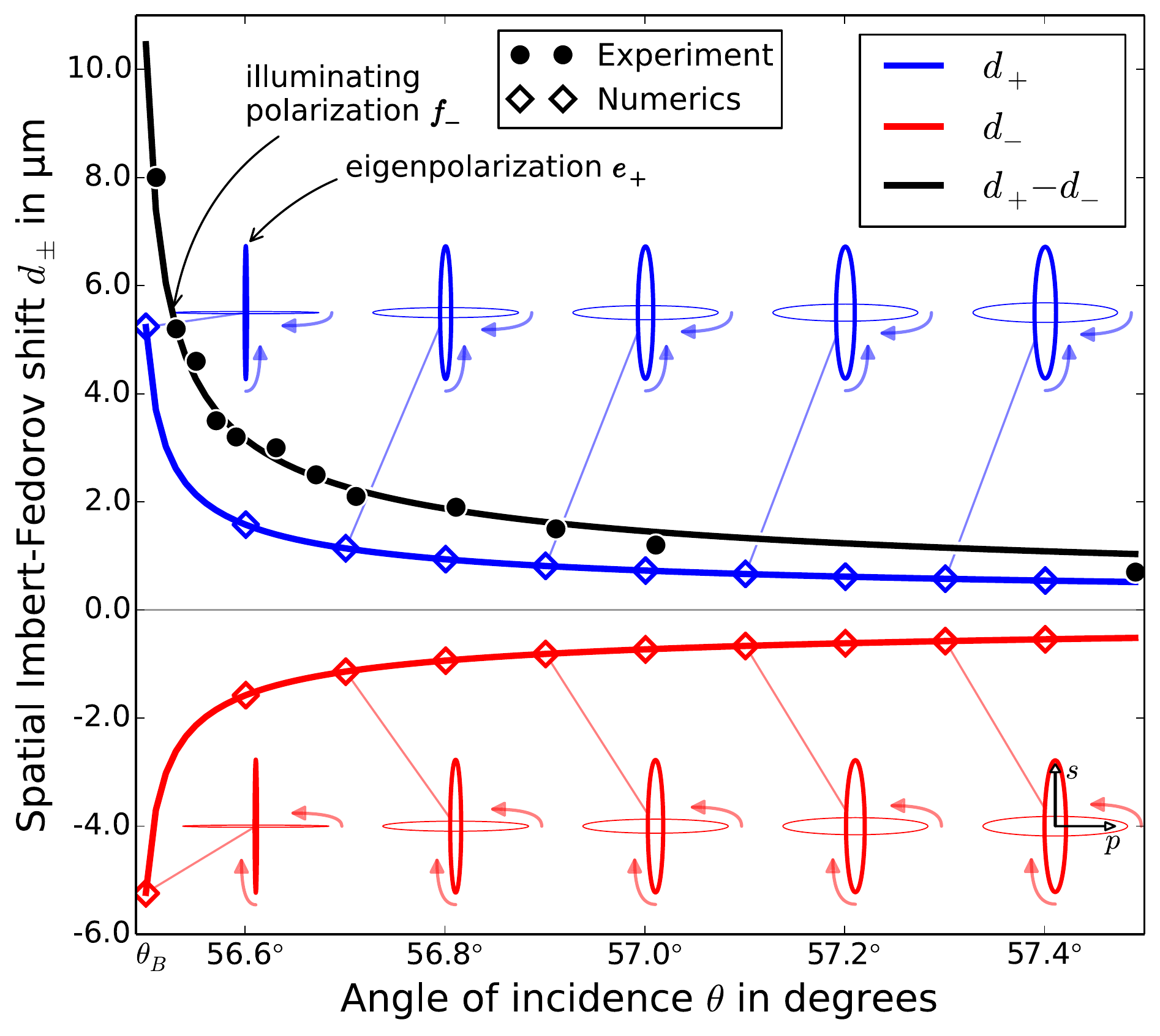}
	\caption{\label{fig:graph} (Color online)
	Measured and calculated shifts: The graph shows the differential spatial IF shifts $d_+ - d_-$ (black) and the individual shifts $d_\pm$ (blue, red) as
	functions of the incidence angle $\theta$ above the Brewster angle $\theta_{\mathrm{B}} = 56.49^\circ$ for $n=1.5103.$ Black dots ($\bullet$) indicate
	experimentally 
	measured values ($d_+ - d_-$), diamonds ($\diamond$) numerically obtained values ($d_\pm$). Solid lines correspond
	to theoretical curves from (\ref{eq:Ayevals}). 
	The plot also shows polarization ellipses for the eigenvectors $\boldsymbol{e}_\pm$ (bold) and the illuminating polarizations $\boldsymbol{f}_\mp$ (thin)
	corresponding to $d_\pm$, which tend to linear on approaching $\theta_{\mathrm{B}}$. 
	The small arrows indicate the handedness of the elliptical polarization.
   }
 \end{center}  
\end{figure}

As the incidence angle approaches $\theta_{\mathrm{B}}$ (from above), the illuminating polarization approaches pure linear $p$-polarization, so most of the incident light is transmitted and not reflected.
This can cause problems if the experimental polarization purity is insufficient, as the unwanted part, orthogonal to the desired eigenpolarization, is reflected much more strongly.
If we denote the intensity extinction ratio with $\gamma$, the light after the polarization preparation stage is in a state
$\boldsymbol{f}_\mp + \sqrt{\gamma} \, \boldsymbol{e}_\pm$, where $\boldsymbol{e}_\pm$ is the orthogonal complement of the illuminating polarization and commonly $\gamma \ll 1.$
As $\overline{\mathbf{R}}$ is not unitary in partial reflection, after reflection the wanted and unwanted polarizations are no longer orthogonal to each other and the polarization is in the state $\sqrt{|r_s||r_p|}\boldsymbol{e}_\pm + \sqrt{\gamma} \, \overline{\mathbf{R}} \cdot \boldsymbol{e}_\pm$. 
For angles close to the Brewster angle, where $r_p \to 0$, the ratio of intensities between the unwanted and desired polarizations is approximately $\gamma |r_s|/|r_p|$. 
For our choice of wavelength and reflection from the BK7 interface we experience a ratio of $|r_s|/|r_p| \approx 1800$ (at $\theta=56.51^\circ$), accounting for why we cannot use commercial liquid-crystal based tunable waveplates for the polarization control.
Instead we use the combination of polarizer (extinction ratio 1:10000) and conventional waveplates, for which we measure a combined extinction ratio of 1:7000, and  which is sufficient for our purposes.

In the absence of a stable reference position we perform a differential measurement between $d_+$ and $d_-.$
As the beam position is measured for the two polarizations $\boldsymbol{f}_\mp$ (or $\boldsymbol{e}_\pm$) at different times consecutively, we need to correct for unavoidable mechanical drifts; this is done via multiple measurements with alternating polarization states.
Fig.~\ref{fig:graph} shows the measured differential shift $d_+ - d_-$ as function of the incidence angle $\theta$.
We have also performed numerical calculations to verify our analytical expressions for the individual shifts $d_\pm$.
The angle dependent eigenpolarizations $\boldsymbol{e}_\pm$ as well as the illuminating polarizations $\boldsymbol{f}_\pm$ are plotted in Fig.~\ref{fig:graph} as blue and red ellipses with arrows indicating the handedness.

The existence of real eigenvalues (such as the large observed shifts $d_\pm$) for non-Hermitian operators is often associated with $\mathcal{PT}$ symmetry \cite{BenderBoettcher:PRL80:1998}.
In the absence of a Schr\"odinger-type equation invariant under combined time and parity inversion, we note that the operator $\mathbf{A}_y$ is a purely imaginary superposition of the Pauli matrices $\boldsymbol{\sigma}_1$ and $\boldsymbol{\sigma}_2$ and falls into the category of general $\mathcal{PT}$-symmetric matrices as defined by \cite{Wang+:JPA43:2010}.
Physically, such systems are often prone to sudden changes connected to the onset or breakdown of $\mathcal{PT}$ symmetry; the Brewster angle marks such an onset owing to fact the determinant of $\overline{\mathbf{R}}$ passes through zero and the associated flip of sign for the reflected, $p$-polarized polarization component.

\begin{figure}
  \begin{center}
   \includegraphics[width=\linewidth]{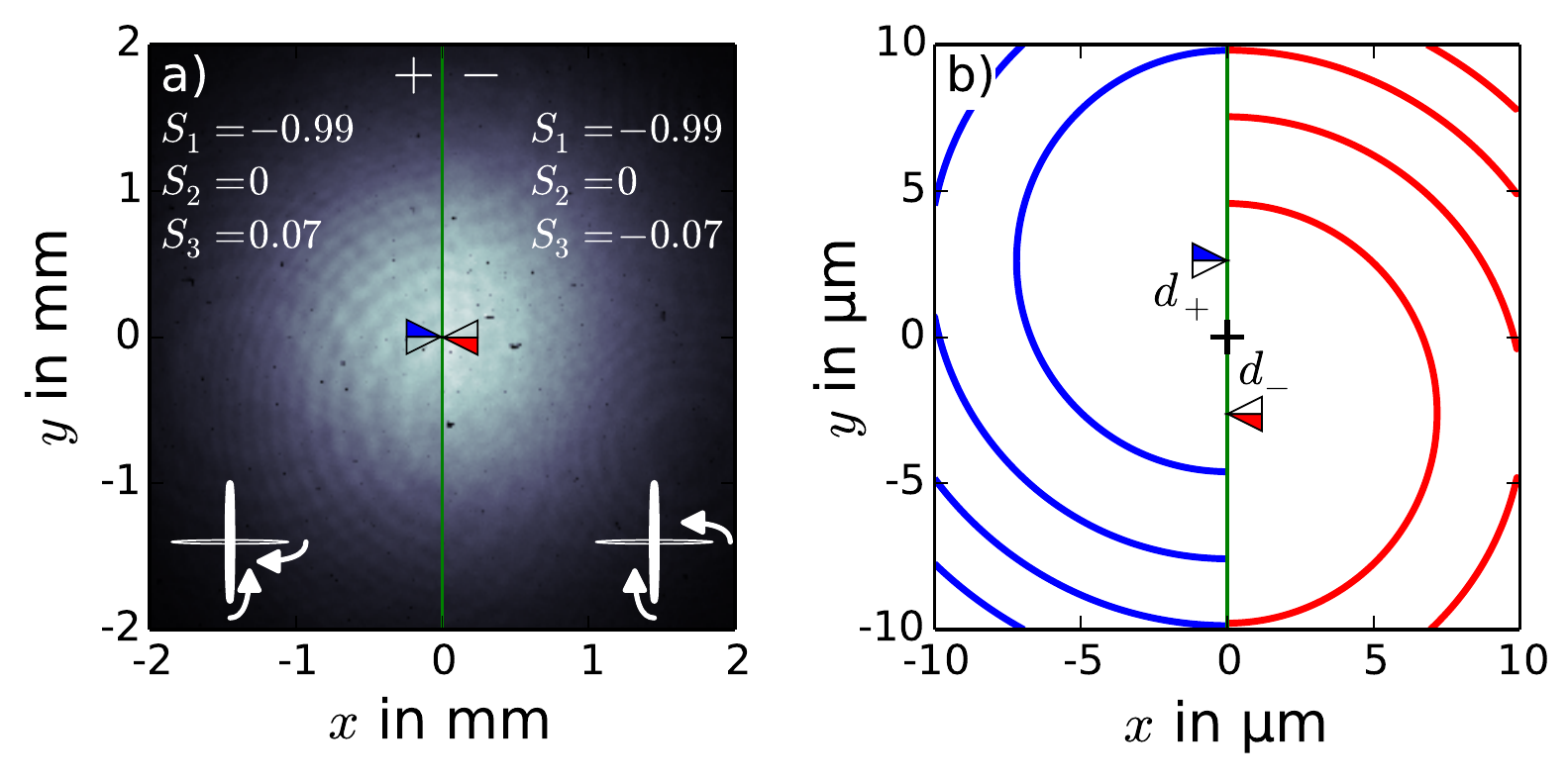}
	\caption{\label{fig:beamprofile} (Color online)
	Measured and calculated beam profiles for $\theta=56.53^\circ$: Split screen comparison of the beam profiles after reflection corresponding to the eigenpolarizations $\boldsymbol{e}_+$ (left panel) and 
	$\boldsymbol{e}_-$ (right panel). 
	a) shows experimental beam profiles recorded with a beam profiler. We also give the Stokes parameter of $\boldsymbol{e}_\pm$ and the polarization ellipses for both $\boldsymbol{e}_\pm$ and the illuminating eigenpolarizations $\boldsymbol{f}_\mp.$
	b) shows numerically calculated beam profiles at a magnification of 500 to highlight the displacement of the centre of the beam. 
	This illustrates that the transverse spatial shift does not lead to any beam deformations.
   }
 \end{center}  
\end{figure}

The measurements of the transverse shift reported here are large, but the magnitude is fundamentally not due to any weak enhancement by means of a weak value measurement, which entails postselection with an analyzer \cite{Luo+:PRA84:2011b,Kong+:APL100:2012}.
It is, however, possible to intepret the form of the operator $\mathbf{A}_y$ at the Brewster angle as an inherent postselection close to the Brewster angle as the difference in the two anti-diagonal entries of $\mathbf{A}_y$ diverges \cite{Gorodetski+:PRL109:2012}.
The similarity to a weak value becomes formally apparent if we chose to factor out $\mathbf{\overline{R}}$ to the left in Eq.~(\ref{eq:Rexp}).
In this case the definition of $\mathbf{A}_y$ changes (though the eigenvalues remain the same), and the IF shift is no longer given by the expectation value of $\mathbf{A}_y$ for the state $\boldsymbol{E}_{\mathrm{i}}$, but rather by $(\boldsymbol{E}_{\mathrm{r}}^*\cdot \mathbf{A}_y \cdot \boldsymbol{E})/(\boldsymbol{E}_{\mathrm{r}}^* \cdot \boldsymbol{E}),$ which resembles a weak value of $\mathbf{A}_y$ for a initial polarization $\boldsymbol{E}$ postselected by $\boldsymbol{E}_{\mathrm{r}}=\overline{\mathbf{R}}^T\cdot\overline{\mathbf{R}}\cdot\boldsymbol{E}$ \footnote{See Supplemental Material [http://link.aps.org/\allowbreak supplemental/10.1103/PhysRevLett???.??????], which inlcudes Refs. \cite{Ritchie+:PRL66:1991,Kofman+:PhysRep520:2012}}.

Regardless of the incidence angle, postselection can lead to deformations of the reflected beam from its typically initial Gaussian profile \cite{GoetteDennis:OL38:2013}. 
This effect is distinct from deformations which arise because a significant part of the angular spectrum straddles the Brewster angle \cite{Merano+:NatPhotonics3:2009,Aiello+:OL34:2009}, although both effects may also be combined \cite{Pan+:APL103:2013}.
In our case, without postselection, we expect the profile of the reflected beam to remain Gaussian arbitrarily close to the Brewster angle as long as $\theta - \theta_{\mathrm{B}}$ is larger than the angular spread of the beam.
We therefore test the reflected beams with the giant IF shifts for possible beam deformations by measuring and calculating the reflected intensity profile for an incident angle of $\theta=56.53^\circ=\theta_{\mathrm{B}}+0.04^\circ$.
As we can see from Fig.~\ref{fig:beamprofile} there are no recognizable beam deformations, which confirms that the giant IF effect reported here is not based upon postselection in contrast to previous works \cite{Luo+:PRA84:2011b,Kong+:APL100:2012,Pan+:APL103:2013}.

We have found and confirmed experimentally the existence of a transverse spatial beam shift at Brewster incidence that is very similar in form and magnitude to the well-known longitudinal Goos-H\"anchen shift at critical incidence. 
The reason for this similarity is an analogous change in the symmetry properties of the operators responsible for the respective beam shifts; this demonstrates that there is a closer relation between longitudinal and transverse shifts than previously anticipated.

The eigenpolarization concept we develop here applies to {\em all} beam shift phenomena, not simply to reflection close to the Brewster angle.
In fact, as indicated originally by Fedorov \cite{Fedorov:DAN105:1955}, the IF shift even in the total reflection regime is maximized by appropriate choice of eigenpolarization (\ref{eq:evect}), which is some incidence angle-dependent, equal weighting of $s$ and $p$ polarizations, but is not circular as usually considered \cite{BliokhBliokh:PRL96:2006}.
As previously emphasized, this does not require postselection (as with enhancements from weak measurement), simply it maximizes within the spectral range of the original Artmann operator.

\begin{acknowledgments}
We would like to thank Konstantin Bliokh for insightful discussion. MRD and JBG acknowledge financial support from the Royal Society and WL from the NWO.
\end{acknowledgments}

\bibliography{IFenhancement}

\pagebreak

\section*{Supplementary information}

We report on a large, spatial Imbert-Fedorov (IF) shift near the Brewster angle without postselection, that is without the help of a polarization analyzer through which the reflected beam is passed before recording its position.
Postselection has the potential to enhance beams shifts greatly, in particular for polarization analyzers that are almost crossed with the on-axis polarization of the reflected beam \cite{Ritchie+:PRL66:1991,HostenKwiat:Science319:2008}; the plane waves which contribute to this postselected beam therefore deviate considerably from the centre of the full reflected beam, making the overall postselected intensity very small.
As a consequence the intensity profile of the postselected beam is usually no longer Gaussian.
Instead, regions of (relatively) high intensity can be found away from the beam axis, which is why every displacement, defined as intensity mean, is amplified.
In our work, however, we do not have such deformations and we decidedly do not use a polarization analyzer to perform a postselection.
Nevertheless the transverse shifts are very large owing to the properties of the Artmann operator $\mathbf{A}_y$ and the special preparation of the light beam, which results in $\mathbf{A}_y$ being acted on by its eigenvector.

In \cite{Gorodetski+:PRL109:2012} the term term `built-in' postselection has been introduced for a plasmonic spin Hall effect which selects, owing to the properties of surface plasmon polaritons, a single polarization.
Operators with 'built-in' postselection can show the characteristic behaviour of weak values going throug cross-polarization \cite{DiLorenzo:PRA85:2012,Kofman+:PhysRep520:2012}, without a separate postselection by a polarization analyzer taking place.
The transverse Artmann operator 
\begin{equation}
	\mathbf{A}_y = \rmi \frac{\cot\theta}{k} \left( \begin{smallmatrix} 0 & -(1 + r_p / r_s) \\ (1 + r_s/r_p) & 0 \end{smallmatrix} \right)
\end{equation}
features a similar selection of a single polarization close to the Brewster angle, as the diagonal entry $(1 + r_s/r_p)$ diverges when $\theta \to \theta_B$ (from above).
This leads to a final polarization state tending to pure linear $p$ polarization in the same limit.
Of course, $\mathbf{A}_y$, is the only the first term of a systematic expansion and a more careful analysis requires a regularization by higher order terms \cite{GoetteDennis:OL38:2013}.

It is possible to highlight the connection to weak values by opting for a different choice of factorizsation in our approximation (1) of the main article. 
Factoring $\overline{\mathbf{R}}$ out to the right, as in the main article, allows us to present the shift as expectation value of the Artmann operator. 
If instead we factor $\overline{\mathbf{R}}$ to the left
\begin{eqnarray}
   \mathbf{R}(\boldsymbol{k}) & = & \overline{\mathbf{R}} + k_x \mathbf{R}_x + k_y \mathbf{R}_y + O(K^2), 
    \\
   & = & \overline{\mathbf{R}}(\mathbf{1} + \rmi k_x \tilde{\mathbf{A}}_x + \rmi k_y \tilde{\mathbf{A}}_y) + O(K^2), 
    \\
   & = & \overline{\mathbf{R}}\exp(\rmi [k_x \tilde{\mathbf{A}}_x + k_y \tilde{\mathbf{A}}_y]) + O(K^2),
\end{eqnarray}
where $\boldsymbol{K} = (k_x, k_y)$ is the transverse wavevector, we end up with a slightly different definition for the Artmann operators:
\begin{equation}
\begin{array}{rclcl}	
   \tilde{\mathbf{A}}_x  & = & \frac{\rmi}{k} \left(\begin{smallmatrix} r_p'/r_p & 0 \\ 0 & r_s'/r_s \end{smallmatrix}\right) & = & \mathbf{A}_x, \\
   \tilde{\mathbf{A}}_y  & = & \rmi \frac{\cot\theta}{k} \left( \begin{smallmatrix} 0 & (1 + r_s/r_p) \\ -(1 + r_p / r_s) & 0 \end{smallmatrix} \right) & = & \mathbf{A}_y^T,
\end{array}   
\end{equation}
where $\mathbf{A}_y^T$ is the transpose (without conjugation) of $\mathbf{A}_y$,
The eigenvalues remain unchanged, but the left and right eigenpolarization are interchanged such that $\tilde{\boldsymbol{e}}_\pm = \boldsymbol{f}_\mp$ and $\tilde{\boldsymbol{f}}_\pm = \boldsymbol{e}_\mp$.
Interestingly, the distinction between (right) eigenpolarization and illuminating polarization is not necessary in this case, as $\tilde{\mathbf{A}}_y$ acts directly on a polarization state. 
However, the most important change on factoring $\overline{\mathbf{R}}$ to the left, is that the shift can no longer be interpreted as the expectation value $k^{-1}\boldsymbol{E}_i^*\cdot\mathbf{A}_y\cdot\boldsymbol{E}_i$ of $\mathbf{A}_y$, but rather by $k^{-1}\boldsymbol{E}_r^*\cdot\tilde{\mathbf{A}}_y\cdot\boldsymbol{E}/\boldsymbol{E}_r^*\cdot\boldsymbol{E}$, which resembles a weak value of $\tilde{\mathbf{A}}_y$ for the polarization $\boldsymbol{E}$ postselected by $\boldsymbol{E}_r = \overline{\mathbf{R}}^T\cdot\overline{\mathbf{R}}\cdot\boldsymbol{E}.$

\end{document}